\documentclass[12pt]{amsart}
\begin{document}
\setcounter{tocdepth}{4}
\newtheorem{theorem}{Theorem}[section]
\newtheorem{lemma}{Lemma}[theorem]
\newtheorem{cor}{Corollary}[theorem]
\newtheorem{prop}[theorem]{Proposition}
\theoremstyle{definition}
\newtheorem{definition}{Definition}[section]
\newtheorem{example}{Example}[section]
\numberwithin{equation}{section}

\title[Coupled nonlinear Schr\"odinger equations]{Quasi-periodic and periodic solutions for
systems of coupled nonlinear Schr\"odinger equations}

\author{P L Christiansen }
\address{Institute of Mathematical Modeling
\\Danish Technical
University
\\DK-2800 Lyngby }
\email{plc@imm.dtu.dk}

\author{J C Eilbeck }
\address{
Department of Mathematics\\
Heriot-Watt University\\
, Edinburgh \\
UK}
\email{J.C.Eilbeck@ma.hw.ac.uk}

\author{V Z Enolskii }
\address{
Theoretical Physics Division\\
NASU Institute of Magnetism\\
36--b Vernadsky str.\\
Kiev-680\\
252142\\
Ukraine}
\email{vze@imag.kiev.ua}

\author{N A Kostov }
\address {
Institute of Electronics\\
Bulgarian Academy of Sciences\\
Blvd. Tsarigradsko shosse 72, Sofia 1784,
\\Bulgaria}
\email{nakostov@ie.bas.bg}

\thanks{The research described in this publication was supported in part by
grants from the Civil Research Development Foundation, CRDF grant no.
UM1-325, the INTAS grant no. 96-770 (JCE and VZE) and the
Engineering and Physical Sciences Research Committee (JCE and VZE and NAK)}

\maketitle

\begin{abstract}

We consider travelling periodic and quasi-periodic wave solutions of a
set of coupled nonlinear Schr{\"o}dinger equations. In fibre optics these
equations can be used to model single mode fibers with strong
birefingence and two-mode optical fibers. Recently these equations
appear as model, which describe pulse-pulse interaction in
wavelength-division-multiplexed channels of optical fiber transmission systems.
Two phase quasi-periodic solutions for integrable Manakov system are given in
terms of two dimensional Kleinian functions. The reduction of quasi-periodic
solutions to elliptic functions is discussed.
New solutions in terms generalized Hermite polynomials, which are associated
with two-gap Treibich-Verider potentials are found.

\end{abstract}

\section{Introduction}

We consider the system of two coupled nonlinear Schr\"odinger equations
\begin{eqnarray}
&&\imath {\mathcal  U}_t+{\mathcal  U}_{xx}+(\kappa {\mathcal  U}{\mathcal  U}^{\ast}
+\chi {\mathcal  V}{\mathcal  V}^{\ast}){\mathcal  U}=0,  \nonumber \\
&&  \label{manakov} \\
&&\imath {\mathcal  V}_t+{\mathcal  V}_{xx}
+( \chi{\mathcal  U}{\mathcal  U}^{\ast}+ \rho {\mathcal  V}{\mathcal  V}^{\ast}){\mathcal  V}=0,
\nonumber
\end{eqnarray}
where $\kappa,\chi,\rho$ are some constants. The integrability of this system
is proven by Manakov
\cite{ma74} only for the case $\kappa=\chi=\rho$, which we shall refer
as {\it Manakov system}.

The equations (\ref{manakov})  are important for a number of physical applications when
$\chi$ is positive and all remaining constants equals to 1.
For example for two-mode optical fibers $\chi=2$
 \cite{ccp82} and for propagation of two modes in fibers with
strong birefringence $\chi=\frac{2}{3}$ \cite{me87} and in
the general case $\frac{2}{3}\leq \chi \leq 2$ for elliptical
eigenmodes. The special value $\chi=1$ (Manakov system)
corresponds to at least two possible cases, namelly the case of a purely
electrostrictive nonlinearity or, in the elliptical birefringence,
when angle between the major and minor axes of the birefingence
ellipse is approximately $35^{o}$. The experimental observation of Manakov
solitons in crystals is reported in \cite{ksaa96}. Recently Manakov
model appear in the Kerr-type approximation of photorefractive crystals
\cite{kpsv98}. The pulse-pulse collision between
wavelength-division-multiplexed channels of optical fiber transmission systems
are described with equations (\ref{manakov}) $\chi=2$, \cite{meg91, kmw96,
hk95,ko97}.

General quasi-periodic solutions in terms of $n$-phase theta functions for
integrable Manakov system are derived in \cite{ahh93}, while a series of
special solutions are given in \cite{allss95,puc98,phf98,pp99}. The authors of
this paper discussed already quasi-periodic and periodic solutions associated
to Lam\'e and Treibich-Verdier potentials for nonintegrable system of coupled
nonlinear  Schr\"odinger
equations in frames of a special ansatz \cite{ceek95}. We also mention the
method of constructing elliptic finite-gap solutions of the stationary KdV
and AKNS hierarchy, based on a theorem  due to Picard, is proposed in
\cite{gw96,gw98b,gw98a} and the method developed by Smirnov in series of
publications, the
review paper\cite{sm94} and \cite{sm97,sm97a}.

In the present paper we investigate integrable Manakov system being restricted
to the system integrable in terms of ultraelliptic functions by
introducing special ansats, which was recently apllied by {\it Porubov and
Parker}  \cite{pp99} to analyse special classes of elliptic solutions of the
Manakov system $(\kappa=\chi=\rho=1)$.  More precisely, we seek solution of (\ref{manakov})
in the form

\begin{eqnarray}
{\mathcal  U}(x,t)=q_1(x) \,\mathrm{ exp}\left \{\imath a_1 t+\imath C_1\int\limits_{\cdot}^x
 \frac{{\mathrm d}x}{q_1^2(x)}\right\},\label{ansatz}\\
{\mathcal  V}(x,t)=q_2(x) \,\mathrm{ exp}\left \{\imath a_2 t+\imath C_2\int\limits_{\cdot}^x
 \frac{{\mathrm d}x}{q_2^2(x)}\right\},\nonumber
\end{eqnarray}
where the functions  $q_{1,2}(x)$ are supposed to be real and $a_1,a_2,C_1,C_2$ are
real constants. Substituting (\ref{ansatz}) into
(\ref{manakov}) we reduce the system to the equations
\begin{eqnarray}
\frac{\partial^2 q_1 }{\partial x^2} +\rho q_1^3+\chi q_1q_2^2-a_1q_1-\frac{C_1^2}{q_1^3}=0
\label{system2}\\
\frac{\partial^2 q_2 }{\partial x^2} +\kappa q_2^3+\chi q_2q_1^2-a_2q_2-\frac{C_2^2}{q_2^3}=0.
\nonumber
\end{eqnarray}
The system  (\ref{system2}) is the natural hamiltonian two-particle system with the
hamiltonian of the form
\begin{eqnarray}
H&=& \frac12p_1^2+ \frac12p_2^2+\frac14( \rho q_1^4+2\chi q_1^2q_2^2+\kappa q_2^4)
\cr&-&\frac12 a_1q_1^2
-\frac12a_2q_2^2+\frac12\frac{C_1^2}{q_1^2}+\frac12\frac{C_2^2}{q_2^2},
\end{eqnarray}
where $p_1(t)= {\mathrm d}q_i(t)/dt$.

These equations describe the motion of particles interacting with the
quartic potential $Aq_1^4+Bq_1^2q_2^2+Cq_2^4$ and perturbed
by inverse squared potential. Nowdays four nontrivial cases of complete
integrabilty are known for nonperturbed potential
(i) A:B:C= 1:2:1,
(ii)  A:B:C= 1:12:16, (iii)  A:B:C= 1:6:1,(iv)  A:B:C= 1:6:8. Cases (i), (ii)
and (iii) are separable in respectively ellipsoidal, paraboloidal and Cartesian
coordinates, while the case (iv) is separable in general sence \cite{rrg94}.
The cases (ii)  appears as one of the entries to polynomial hierarchy discussed in \cite{eekl93aa}
the cases  (iii) and (iv) are proved to be canonically equivalent under the
action of Miura map restricted to the stationary of coupled KdV systems associated
with fourth order Lax operator\cite{bef95}. Moreover all the cases
(i)-(iv) permit the deformation
of the potential by linear conbination of inverse squares and squares with
certain limitations on the coefficients
\cite{eekl93aa,bef95}. There are also known Lax representations for all
these cases which yield hyperelliptic  algebraic curves in the cases (i) and (ii) and
4-gonal curve in the cases (iii) and (iv).

Although each from the system enumerated yield nontrivial classes of solutions  of
the system (\ref{manakov})  we shall discuss
further only the case (i). The integrability of this case and separability
in ellipsoidal coordinates was proved by
{\it  Wojciechowski} \cite{w85}   (see also
\cite{k89,t95}). We employ this result to integrate the
system in terms of ultraelliptic functions (hyperelliptic functions of the
genus two curve) and then execute reduction of hyperelliptic functions to
elliptic ones by imposing additional constrains on the parameters of the system.

The paper is organised as follows. In the first section we construct the Lax
representation of the system, develop the genus two algebraic curve, which is
associated to the system and reduce the problem to solution of the Jacobi
inversion problem associated with genus two algebraic curve. In the section
two develop the integration of the system in terms of {\it Kleinian hyperelliptic
functions}  which represent a natural generalization of Weierstrass elliptic
functions to hyperelliptic curved of hidher genera; recently this realization
of abelian functions was discussed in \cite{bel97b,bel97c,eel99}. We explain
in the section the outline of the Kleinian realization of hyperelliptic functions
and give the principle formulae for the case of genus two curve. In Section 4
we develop reduction of Kleinian hyperelliptic function to elliptic functions
in terms of {\it Darboux coordinates} for the curve admiting additional
involution. In this way a quasiperodic solution in terms of elliptic functions
is obtained. In the last section we construct a set of elliptic periodic
solutions on the basis of application of spectral theory for the Hill equation
with elliptic potential.

\section{Lax representation}

The system $1:2:1$  $(\kappa=\chi=\rho=1)$  is a completely integrable hamiltonian
system

\begin{eqnarray}
\frac{\partial^2 q_1 }{\partial x^2} +(q_1^2+q_2^2)q_1-a_1q_1-\frac{C_1^2}{q_1^3}=0,\cr
\label{system}\\
\frac{\partial^2 q_2 }{\partial x^2} +(q_1^2+q_2^2)q_2-a_2q_2-\frac{C_2^2}{q_2^3}=0
\nonumber
\end{eqnarray}
with the Hamiltonian
\begin{equation}
H=\frac12\sum_{i=1}^2
p_i^2+\frac14\left(q_1^{2}+q_2^{2}\right)^2-\frac12a_1 q_1^2-
\frac12 a_2 q_2^2+ \frac12\frac{C_1^2}{q_1^2}+
\frac12\frac{C_2^2}{q_2^2}, \label{H}\end{equation} where
the variables $(q_1,p_1;q_{2},p_{2})$
are the canonicaly conjugated variables with respect to the standard Poisson
bracket, $\{\cdot\;;\;\cdot\}$.

This system permits the Lax representation (special case of Lax
representation given in \cite{k98}).

\begin{eqnarray}
\frac{\partial L(\lambda)}{\partial \zeta}&=&[M(\lambda),L(\lambda)],\cr
\quad
L(\lambda)&=&\left(
\begin{array}{cc}
V(\lambda) & U(\lambda) \\
W(\lambda) & -V(\lambda)
\end{array}
\right),\quad M=\left(
\begin{array}{cc}
0 & 1 \\
Q(\lambda) & 0
\end{array}
\right)  \label{lax}
\end{eqnarray}
is equvalent to the (\ref{system}), where $U(\lambda),W(\lambda),Q(\lambda)$ have the form

\begin{eqnarray*}
U(\lambda)&=&-a(\lambda)\left(1+\frac{1}{2}\frac{q_1^2}{\lambda-a_1}
+\frac{1}{2}\frac{q_2^2}{\lambda-a_2}\right),  \label{u} \\
V(\lambda)&=&-\frac12\frac{\mathrm{d}} {\mathrm{d\zeta}}
U(\lambda) , \label{v} \\
W(\lambda)&=&a(\lambda)\left(-\lambda+\frac{q_1^2}{2}+\frac{q_2^2}{2}
+\frac12\left(p_1^2+\frac{C_1^2}{
q_1^2}\right)\frac{1}{\lambda-a_1}\right.\cr&+&\left.
\frac12\left(p_2^2+\frac{C_2^2}{q_2^2}\right)
\frac{1}{
\lambda-a_2} \right) , \label{w} \\
Q(\lambda)&=&\lambda-q_1^2-q_2^2,  \label{q}
\end{eqnarray*}
where $a(\lambda)=(\lambda-a_1)(\lambda -a_2)$.

The Lax representation yields hyperelliptic curve
$V=(\nu,\lambda)$, \[ \det(L(\lambda)-\frac12\nu 1_2)=0, \]
where $1_2$ be $2\times2$ unit matrix and  is given explicitly as

\begin{eqnarray}
\nu^2&=&4(\lambda-a_1)(\lambda-a_2)(\lambda^3-\lambda^2(a_1+a_2)
+\lambda(a_1a_2-H)-F)\cr
&-&C_1^2(\lambda-a_2)^2-C_2^2(\lambda-a_1)^2, \label{curve}
\end{eqnarray} where $H$ is the hamiltonian (\ref{H}) and the
second independent integral of motion $F$, $\{F;H\}=0$ is given as
\begin{eqnarray}
F&=&\frac14(p_1q_2-p_2q_1)^2+\frac12(q_1^2+q_2^2)(a_1a_2-\frac12a_2q_1^2
-\frac12a_1q_2^2)\cr
&-&\frac12p_1^2a_2-\frac12p_2^2a_1
-\frac14\frac{(2a_2-q_2^2)C_1^2}{q_1^2}
-\frac14\frac{(2a_1-q_1^2)C_2^2}{q_2^2}. \label{F}\end{eqnarray}
We remark, that the parameters $C_i$ are linked with coordinates
of the points $(a_i,\nu(a_i))$ by the formula
\begin{equation}
C_i^2=-\frac{\nu(a_i)^2}{(a_i-a_j)^2},\quad i,j=1,2.\label{ccc}
\end{equation}

Let us write the curve (\ref{curve}) in the form
\begin{eqnarray}
\nu^2=4\lambda^5+\alpha_4\lambda^4+\alpha_3\lambda^3+\alpha_2\lambda^2
+\alpha_1\lambda+\alpha_0,\label{curvecan}
\end{eqnarray}
where the {\it moduli} of the curve $\alpha_i$ are expressible in
terms of physical parameters - level of energy $H$ and constants
$a_1,a_2$, $C_1,C_2$ as follows
\begin{eqnarray}
\alpha_4&=&-8(a_1+a_2),\cr
\alpha_3&=&-4H+4(a_1+a_2)^2+8a_1a_2,\cr
\alpha_2&=&4H(a_1+a_2)-4F-C_1^2-C_2^2-8a_1a_2(a_1+a_2),\cr
\alpha_1&=&4F(a_1+a_2)-4a_1a_2H+2C_1^2a_2+2C_2^2a_1 +4a_1^2a_2^2,\cr
\alpha_0&=&-4a_1a_2F-C_1^2a_2^2-C_2^2a_1^2.\nonumber
\end{eqnarray}

Let us define new coordinates $\mu_1,\mu_2$ as zeros of the entry $
U(\lambda)$ to the Lax operator. Then
\begin{eqnarray}
q_1^2=2\frac{(a_1-\mu_1)(a_1-\mu_2)}{a_1-a_2},\quad q_2^2=2 \frac{
(a_2-\mu_1)(a_2-\mu_2)}{a_2-a_1}.\label{qcoord}
\end{eqnarray}
The definition of $\mu_1,\mu_2$ in the combination with the Lax representation
comes to the equations
\begin{equation}
\nu_i=V(\mu_i)=-\frac12\frac{\partial}{\partial x}U(\mu_i),\quad i=1,2,
\end{equation}
wich can be transformed to the equations of the
the form\footnote{In what follows we shall denote the integral bounds by the
second
coordinate of the curve $V=V(\nu,\lambda)$. (\ref{curve})}
\begin{eqnarray}
u_1=\int_{a_1}^{\mu_1}\mathrm{d}u_1 +\int_{a_2}^{\mu_2}du_1, \\
u_2=\int_{a_1}^{\mu_1}\mathrm{d}u_2 +\int_{a_2}^{\mu_2}
du_2\label{jip}
\end{eqnarray}
where $\mathrm{d}u_{1,2}$ denote independent canonical holomorphic differentials
\begin{equation}
\mathrm{d}u_1= \frac{\mathrm{d}\lambda}{\nu},\quad \mathrm{d}u_2=\frac{\lambda
\mathrm{d}\lambda}{\nu} .\label{hodbas}
\end{equation}
and $u_1=a,u_2=2x+b$ with the constants $a,b$ defining by the initial
conditions. The integration of the problem is then reduces to the solving of
the {\it Jacobi inversion
problem } associated with the curve, which consist in the expession of the
symmetric functions of $(\mu_1,\mu_2,\nu_1,\nu_2)$ as function of two complex
variables $(u_1,u_2)$.

\section[Kleinian hyperelliptic
functions]{Exact solutions in terms of Kleinian hyperelliptic
functions}

In this section we give the trajectories of the system under
consideration
in terms of Kleinan hyperelliptic functions (see, e.g.
\cite{ba97,bel97c}),
being associated with the algebraic curve of genus two (\ref{curvecan}) which can be also written
in the form
\begin{eqnarray}
\nu^2&=4\prod_{i=0}^{4}(\lambda-\lambda_{i}),  \label{gen2}
\end{eqnarray}
where $\lambda_{i}\neq \lambda_{i}$ are branching points.
At all real branching points the closed intervals
$[\lambda_{2i-1},\lambda_{2i}],i=0,\ldots 4$ will be referred further as lacunae
\cite{zmnp80,mm75}.
Let us equip the curve with a homology basis $({\mathfrak a}_1,{\mathfrak a}_2;
{\mathfrak  b}_1, {\mathfrak b}_2)\in H_1(V,{\mathbb Z})$ and fix the basis in  the space of holomorphic differentials as in
(\ref{hodbas}). The associated canonical meromorphic differentials of the
second kind $\mathrm{ d}\boldsymbol {r}^T=(\mathrm{ d}r_1,{\mathrm
d} r_2)$ have the form
\begin{equation}{\mathrm
d}r_1=\frac{\alpha_3\lambda+2\alpha_4\lambda^2+12\lambda^3}{
4\nu}d\lambda,\qquad {\mathrm d}r_2=\frac{\lambda^2}{ \nu}d\lambda.\label{rr}
\end{equation}
The $2\times 2$ matrices of their periods,
\begin{eqnarray*}
2\omega&=&\left(\oint_{{\mathfrak a}_k}{\mathrm
d} u_l\right)_{k,l=1,2},\quad
2\omega'=\left(\oint_{{\mathfrak b}_k}{\mathrm
d} u_l\right)_{k,l=1,2},\\
2\eta&=&\left(\oint_{{\mathfrak a}_k}{\mathrm
d} r_l\right)_{k,l=1,2},\quad
2\eta'=\left(\oint_{{\mathfrak b}_k}{\mathrm
d} r_l\right)_{k,l=1,2}
\end{eqnarray*}
satisfy the equations,
\[\omega'\omega^T-\omega{\omega'}^T=0,\quad
\eta'\omega^T-\eta{\omega'}^T=-\frac{\imath\pi}{2}1_2,\quad
\eta'\eta^T-\eta{\eta'}^T=0,
\]
which generalizes the Legendre relations between complete elliptic
integrals to the case $g=2$.

The fundamental $\sigma$ function in this case is a natural
generalization of the Weierstrass elliptic $\sigma$ function and
is defined as follows
\begin{eqnarray*}
\sigma(\boldsymbol{u})&=&\frac{\pi}{\sqrt{\mathrm{det}(2\omega)}}
\frac{\epsilon}{\sqrt[4]{\prod_{1\leq i<j\leq 5}(a_i-a_j)}}\\
&\times&\exp\left\{\boldsymbol{
u}^T\eta(2\omega)^{-1}\boldsymbol{u}\right\}
\theta[\varepsilon]((2\omega)^{-1}
\boldsymbol{ u}|\omega'\omega^{-1}),
\end{eqnarray*}
where $\epsilon^8=1$ and $\theta[\varepsilon](\boldsymbol{
v}|\tau)$ is the $\theta$ function with an odd characteristic
$[\varepsilon]=\left[\begin{array}{cc}\varepsilon_1&\varepsilon_2\\
\varepsilon_1'&\varepsilon_2'\end{array}\right]$, which is the characteristic
of the vector of Riemann constants,
\[\theta[\varepsilon](\boldsymbol{ v}|\tau)=\sum_{\boldsymbol{
 m}\in{\mathbb Z}^2}\mathrm{\exp}\;\imath\pi\left\{ (\boldsymbol{
m}+{\boldsymbol\varepsilon})^T\tau (\boldsymbol{
m}+{\boldsymbol \varepsilon})+2 (\boldsymbol{
v}+{\boldsymbol \varepsilon}')^T\tau (\boldsymbol{
m}+{\boldsymbol\varepsilon})\right\}. \]

Alternatively the $\sigma $ function can be defined by  its expansion near
$\boldsymbol{u}=0$
\begin{equation}
\sigma (\boldsymbol{u})=u_1+\frac{1}{24}
\alpha_2u_1^3-\frac{1}{3}u_2^3+o(\boldsymbol{u}^5 )\label{ex}
\end{equation}
and the further terms can be computed with the help of bilinear
differential equation \cite{ba07}.

The $\sigma$-function posses the following periodicity property:
put \[
\boldsymbol{E}(\boldsymbol{ m},\boldsymbol{ m}')=\eta\boldsymbol{
m} +\eta'\boldsymbol{ m}',\quad\text{and}\quad
\boldsymbol{\Omega}(\boldsymbol{ m},\boldsymbol{  m }')
  =\omega\boldsymbol{ m} +\omega' \boldsymbol{ m}', \]  where
$\boldsymbol{ m},\boldsymbol{ m}'\in \mathbb{Z}^{n}$, then
\begin{align*} &\sigma[\varepsilon](\boldsymbol{z}+
2{\boldsymbol\Omega} (\boldsymbol{ m }, \boldsymbol{  m
}'),\omega,\omega') =\mathrm{exp} \big\{ 2\boldsymbol{E}^T
(\boldsymbol{ m},\boldsymbol{ m}') \big({\boldsymbol z}+
\boldsymbol{\Omega}(\boldsymbol{  m }, \boldsymbol{  m
}')\big)\big\}\\
&\times \mathrm{exp} \{ -\pi \imath  {\boldsymbol m
}^T{\boldsymbol m}' -2\pi \imath  {\boldsymbol\varepsilon
}^T{\boldsymbol m}' \}
\sigma[\varepsilon]( {\boldsymbol z},\omega,\omega')
\end{align*}
As modular function the Kleinian $\sigma$-function is
invariant under the transformation of the symplectic group, what
represents the important characteristic feature.

We introduce the Kleinian hyperelliptic functions
as second logarithmic derivatives
\begin{eqnarray*}
\wp_{11}(\boldsymbol{ u})&=&-\frac{\partial^2}{
\partial u_1^2}\mathrm{ ln}\; \sigma(\boldsymbol{
u}),\quad \wp_{12}(\boldsymbol{ u})=-\frac{\partial^2}{\partial
u_1\partial u_2}\mathrm{ ln}\; \sigma(\boldsymbol{ u}),\cr
\wp_{22}(\boldsymbol{ u})&=&-\frac{\partial^2}{\partial u_2^2} \mathrm{
ln}\; \sigma(\boldsymbol{ u}).
\end{eqnarray*}
The multi-index symbols $\wp_{i,j,k}$ etc. are defined as logarithmic
derivatives by the variable $u_i,u_j,u_k$ on the corresponding indices
$i,j,k$ etc.

The principal result of the theory is the formula of Klein, which reads
in the case of genus two as follows
\begin{eqnarray}
&&\sum_{k,l=1}^2\wp_{kl}\left(\int_{\infty}^{\mu} {\mathrm d}{\mathbf u}-
\int_{\infty}^{\mu_1 } {\mathrm d}{\mathbf u}-
\int_{\infty}^{\mu_2 } {\mathrm d}{\mathbf u}\right)\mu^{k-1}\mu_i^{l-1}\cr
&=&\frac{F(\mu,\mu_i)+2\nu\nu_i}{4(\mu-\mu_i)^2}, \quad i=1,2,\label{klein}
\end{eqnarray}
where
\begin{equation}
F(\mu_1,\mu_2)=\sum_{r=0}^2\mu_1^r\mu_2^r[2\alpha_{2r}
+\alpha_{2r+1}(\mu_1+\mu_2)].\label{fx1x2}
\end{equation}
By expanding these equalities in the vicinity of the infinity we obtain
the complete set of the relations for hyperelliptic functions.

The first group of the relations represents the solution of the Jacobi
inversion problem in the form
\begin{equation}
\lambda^2-\wp_{22}(\boldsymbol{u})\lambda
-\wp_{12}(\boldsymbol{u})=0,\label{bolza2}\end{equation} that is,
the pair $(\mu_1,\mu_2)$ is the pair of roots of
 (\ref{bolza2}). So we have
\begin{equation}\wp_{22}(\boldsymbol{u})
=\mu_1+\mu_2,\;\wp_{12}(\boldsymbol{u})=-\mu_1\mu_2.\label{b1}\end{equation}
The corresponding $\nu_i$ is expressed as
\index{$\wp$ function!fundamental relation!for genus two}
 \begin{equation}
\nu_i=\wp_{222}(\boldsymbol{u})\mu_i+\wp_{122}(\boldsymbol{ u}),\quad
 i=1,2. \label{y2}
\end{equation}
The functions $\wp_{22},\wp_{12}$ are called basis functions. The function
$\wp_{11}(\boldsymbol{u})$ can be also espressed as symmetric function of $\mu_1,\mu_2$
and $\nu_1,\nu_2$:
\begin{equation} \wp_{11}(\boldsymbol{u})=
\frac{F(\mu_1,\mu_2)-2\nu_1\nu_2}{4(\mu_1-\mu_2)^2},
\label{b2}
\end{equation}
where  $F(\mu_1,\mu_2)$  is given in (\ref{fx1x2}).

Further from (\ref{y2}) we have
\begin{eqnarray}
\wp_{222}(\boldsymbol{ u})&=&\frac{\nu_1-\nu_2}{ \mu_1-\mu_2},\quad
\wp_{221}(\boldsymbol{ u})=\frac{\mu_1\nu_2-\mu_2\nu_1}{ \mu_1-\mu_2},
\nonumber \\
\wp_{211}(\boldsymbol{ u})&=&-\frac{\mu_1^2\nu_2-\mu_2^2\nu_1}{ \mu_1-\mu_2},
\nonumber \\
\wp_{111}(\boldsymbol{ u})&=&\frac{\nu_2\psi(\mu_1,\mu_2)-\nu_1
\psi(\mu_2,\mu_1)}{ 4(\mu_1-\mu_2)^3},  \label{thirdder}
\end{eqnarray}
where
\begin{eqnarray}
\psi(\mu_1,\mu_2) &=& 4\alpha_0 + \alpha_1(3\mu_1 + \mu_2) +
2\alpha_2\mu_1(\mu_1 +\mu_2) \nonumber \\ &+&  \alpha_3\mu_1^2(\mu_1
+3\mu_2)
+ 4\alpha_4\mu_1^2\mu_2 +4\mu_1^2\mu_2(3\mu_1 +\mu_2). \nonumber
\end{eqnarray}

The next group of the relations, which can be drived by the expanding
of the equations (\ref{klein})
are the pairwise products of the $\wp_{ijk}$ functions being
expressed in terms of $\wp_{22},\wp_{12},\wp_{11}$ and
constants $\alpha_s$ of the defining equation (\ref{gen2}). We
give here only basis equations
 \begin{eqnarray*}
\wp_{222}^2&=4\wp_{22}^3+4\wp_{12}\wp_{22}+\alpha_4\wp_{22}^2+4\wp_{11}
+\alpha_3\wp_{22}+\alpha_2,\\
\wp_{222}\wp_{122}&=4\wp_{12}\wp_{22}^2
+2\wp_{12}^2-2\wp_{11}\wp_{22}+\alpha_4\wp_{12}\wp_{22}
\\
&+\frac12\alpha_3\wp_{12}+\frac12\alpha_1
, \\
\wp_{122}^2&=4\wp_{22}\wp_{12}^2-4\wp_{11}\wp_{12}+
\alpha_4\wp_{12}^2-\alpha_0.
\end{eqnarray*}
All such expressions may be rewritten in the form of an {\it extended
  cubic relation} as follows. For arbitrary $\boldsymbol{
  l},\boldsymbol { k}\in {\mathbb C}^4$ the following formula is valid
\cite{ba07}
\begin{equation}\boldsymbol{l}^T\pi\pi^T
\boldsymbol{k}=-\frac14{\det}\;\left(\begin{array}{cc}H&\boldsymbol{l}\\
\boldsymbol{k}^T&0\end{array}\right),\label{kum1}\end{equation}
where
$\pi^T=(\wp_{222},-\wp_{221},\wp_{211},-\wp_{111})$
and   $H$ is the $4\times4$ matrix:
\begin{equation}
H=
 \left (  \begin{array}{cccc}  \alpha_0&  \frac{1}{2}
 \alpha_1&-2  \wp_{11}&-2  \wp_{12}  \\
  \frac{1}{2}  \alpha_1&  \alpha_2+4  \wp_{11}&  \frac{1}{2}
\alpha_3+ 2  \wp_{12}&-2  \wp_{22}  \\-2  \wp_{11}&  \frac{1}{2}
 \alpha_3+2  \wp_{12}&  \alpha_4+4  \wp_{22}&2  \\
-2  \wp_{12}&-2  \wp_{22}&2&0  \end{array}  \right) .
 \end{equation}
The vector $\pi$ satisfies the equation $H\pi=0$,
and so the functions $\wp_{22},\wp_{12}$ and $\wp_{11}$ are
 related by the equation
\begin{equation}\mathrm{ det}\;
H=0.\label{kum5} \end{equation}
The equation (\ref{kum5}) defines
the quartic Kummer surface $\mathbb K$ in $\mathbb C^3$
\cite{hu05}.

The next group of the equations, which is derived as the result of
expansion of the equalities
(\ref{klein}) are the expressions of four index symbols  $\wp_{ijkl}$ as quadrics in $\wp_{ij}$
\begin{eqnarray}
&\wp_{2222}=6\wp_{22}^2+\frac12\alpha_3+\alpha_4\wp_{22}
+4\wp_{12},\label{eeq1}\\
&\wp_{2221}=6\wp_{22}\wp_{12}+\alpha_4\wp_{12}-2\wp_{11},\label{eeq3}\\
&\wp_{2211}=2\wp_{22}\wp_{11}+4\wp_{12}^2+
\frac12\alpha_3\wp_{12}.\label{eeq5}\\
&\wp_{2111}=6\wp_{12}\wp_{11}+\alpha_2\wp_{12}
-\frac12\alpha_1\wp_{22}-\alpha_0,\label{eeq4}\\
&\wp_{1111}=6\wp_{11}^2-3\alpha_0\wp_{22}
+\alpha_1\wp_{12}+\alpha_2\wp_{11}-
\frac12\alpha_0\alpha_4+\frac18\alpha_1\alpha_3.\label{eeq2}
\end{eqnarray}

These equations can be identified with completely integrable
partial differential equations and dynamical systems, which are
solved in terms of Abelian functions of hyperelliptic curve of
genus two. In particular, the first two equations represent the KdV hierarhy with ``times"
$(t_1,t_2)=(u_2,u_1)=(x,t)$,
\begin{equation}{\mathcal  X}_{k+1}[{\mathsf  U}]={\mathcal  R}{\mathcal
X}_{k}[{\mathsf  U}] \end{equation}
where ${\mathcal  R}=\partial_x^2- {\mathsf  U}+c
-\frac12{\mathsf  U}_x\partial^{-1}$, $c=\alpha_4/12$ is the Lenard recursion
operator. The first two equations from the hierarchy are
\begin{equation}
{\mathsf  U}_{t_1}={\mathsf  U}_{x},\quad {\mathsf
U}_{t_2}=\frac12({\mathsf  U}_{xxx}-6{\mathsf  U}_{x}{\mathsf  U}),
\label{kdv}
\end{equation}
the second equation is the KdV equation, which is obtained from (\ref{eeq1}) as the result of
differentiation by $x=u_2$ and setting ${\mathsf  U}=2\wp_{22}+\alpha_4/6$. The
equation (\ref{eeq1})
plays role of the stationary equation in the hierarchy and is obtained as the result of action of
the recursion operator.

Let us introduce finaly the {\it Baker-Akhiezer} function, which in the frames of the formalizm
developed is expressible in terms of the Kleinian $\sigma$-function as follows
\begin{equation}
\Psi(\lambda,\boldsymbol{u})=
\frac{  \sigma\left(\int_{\infty}^{\lambda}{\mathrm d} \boldsymbol{ u}-
{\mathbf u} \right)}  {\sigma(\boldsymbol{u}) }  \mathrm {exp}\left\{
\int_{\infty}^{\lambda}{\mathrm d} {\mathbf r}^T \boldsymbol{u} \right\},\label{BAF}
\end{equation}
where $\lambda$ is arbitrary and $\boldsymbol  u$ is the Abel image of arbitrary
point    $(\nu_1,\mu_1)\times (\nu_2,\mu_2)\in V \times V $. It is straighforward to show
by the direct calculation, being bases on the usage of the relations for three and four-index Kleinian
$\wp$--functions that $\Psi(\lambda,\boldsymbol{u})$ satisfy to the Schr\"odinger equation
\begin{equation}
(\frac{\partial^2}{{\partial u_2}^2}-2\wp_{22}(\boldsymbol{u}))
\Psi(\lambda,\boldsymbol{u})=
\left(\lambda+\frac14\alpha_{4}\right)\Psi(\lambda,\boldsymbol{u})\label{sch}
\end{equation}
for all $(\nu,\mu)$.

Now we are in position to write the solution of the of the system
in terms of Kleinian $\sigma$-functions and identify the constants
in terms of the moduli of the curve. Using (\ref{b1}),(\ref{qcoord}) the solutions
of (\ref{system}) have the
following form
in terms of Kleinian functions $\wp_{22}(\boldsymbol{ u}), \wp_{12}(\boldsymbol{ u})$
\begin{eqnarray}
q_1^2&=&2\frac{a_1^{2}-\wp_{22}(\boldsymbol  u)a_1-\wp_{12}(\boldsymbol  u)}{a_1-a_2},  \cr
q_2^2&=&2\frac{a_2^{2}-\wp_{22}(\boldsymbol  u)a_2-\wp_{12}(\boldsymbol  u)}{a_2-a_1},
\label{solution}
\end{eqnarray}
where the vector $\boldsymbol {u}^T=(a,2x+b)$.

\section[Quasi-periodic elliptic solutions]
{Periodic solutions expressed in terms of elliptic functions of
different moduli}

We consider in this section the reduction Jacobi
(see e.q.\cite{kr03} ) of hyperelliptic integrals to elliptic ones, when
the hyperelliptic curve $V$ has the form
\begin{equation}
w^2=z(z -1)(z -\alpha )(z -\beta )( z -\alpha \beta )
 \label{curver}
\end{equation}

The curve (\ref{curver}) covers two-sheetedly two tori $$\pi _{\pm
}:V=(w,z)\rightarrow E_{\pm }=(\eta _{\pm },\xi _{\pm }),$$
\begin{equation}
\eta _{\pm }^2=\xi _{\pm }(1-\xi _{\pm })(1-k_{\pm }^2\xi _{\pm })
\end{equation}
with Jacobi moduli
\begin{equation}
k_{\pm }^2=-\frac{(\sqrt{\alpha }\mp \sqrt{\beta })^2}{(1-\alpha
)(1-\beta )}
,  \label{jacmod}
\end{equation}
The covers $\pi _{\pm }$ are described by the formulae
\begin{eqnarray}
\eta _{\pm }=-\sqrt{(1-\alpha )(1-\beta )}\frac{z\mp \sqrt{\alpha
\beta }}{
(z-\alpha )^2(z-\beta )^2}w,  \label{r1} \\
\xi=\xi _{\pm}=\frac{(1-\alpha )(1-\beta )z}{(z-\alpha )(z-\beta
)}.  \label{r2} \end{eqnarray} The following formula is valid for
the reduction of holomorphic hyperelliptic differential to the
elliptic ones:  \begin{equation} \frac{d\xi _{\pm }}{\eta _{\pm
}}=-\sqrt{(1-\alpha )(1-\beta )}(z\mp \sqrt{ \alpha \beta
})\frac{\mathrm{d} z}{w}.  \label{r3} \end{equation}

Suppose that the spectral curve (\ref{curvecan}) admits the symmetry of
the (\ref{curver}) and apply the discussed reduction case to the problem.
Then the equations of the Jacobi inversion problem
(\ref{jip}) can be rewritten in the form

\begin{eqnarray}
\sqrt{(1-\beta)(1-\alpha)}\sum_{i=1}^2\int_{z_0}^{z_i}
(z -\sqrt{\alpha\beta})\frac{\mathrm{d}z}
{w}=2u_{+}, \label{j11} \\
\sqrt{(1-\beta)(1-\alpha)}\sum_{i=1}^2\int_{x_0}^{z_i}
(z +\sqrt{\alpha\beta})\frac{\mathrm{d}z}
{w}=2u_{-} .\label{j22}
\end{eqnarray}
with $(\nu_i,\mu_i)=(2w_i,z_i)$ and
\begin{equation}
u_{\pm }=-\sqrt{(1-\alpha )(1-\beta )}(u_2\mp \sqrt{\alpha \beta }u_1)
\end{equation}

Reduce in (\ref{j11},\ref{j22}) hyperelliptic integrals to elliptic ones
according to (\ref{r1},\ref{r2}).
\begin{eqnarray*}
\int_{0}^{\sqrt{\xi(\mu_1)}}\frac{\mathrm{d}x}{\sqrt{(1-x^2)(1-k^2_{\pm}x^2)}}
+\int_{0}^{\sqrt{\xi(\mu_2)}}\frac{\mathrm{d}x}
{\sqrt{(1-x^2)(1-k^2_{\pm}x^2)}}
=u_{\pm},
\end{eqnarray*}

One can further express the symmetric functions of $\mu_1,\mu_2, \nu_1, \nu_2$ on
$V\times V$ in term of elliptic functions of tori $E_{\pm}$. To
this end we introduce the {\it Darboux coordinates} (see
\cite{hu05}, p.105 ) \begin{eqnarray}
X_1=\mbox{sn}(u_{+},k_{+})\mbox{sn}(u_{-},k_{-}),\cr
X_2=\mbox{cn}(u_{+},k_{+})\mbox{cn}(u_{-},k_{-}),  \label{darb}
\\ X_3=\mbox{dn}(u_{+},k_{+})\mbox{dn}(u_{-},k_{-}), \nonumber
\end{eqnarray}
where $\mbox{sn}(u_{\pm },k_{\pm }),\mbox{cn}(u_{\pm },k_{\pm }),\mbox{dn
}(u_{\pm },k_{\pm })$ are standard Jacobi elliptic functions.

We apply further the addition theorem for Jacobi
elliptic functions,
\begin{eqnarray}
\mbox{sn}(u_1+u_2,k)=\frac{s_1^2-s_2^2}{s_1c_2d_2-s_2c_1d_1},\cr
\mbox{cn
}(u_1+u_2,k)=\frac{s_1c_1d_2-s_2c_2d_1}{s_1c_2d_2-s_2c_1d_1},\cr
\mbox{dn}
(u_1+u_2,k)=\frac{s_1d_1c_2-s_2d_2s_1}{s_1c_2d_2-s_2c_1d_1},  \nonumber
\end{eqnarray}
where we denoted $s_i=\mbox{sn}(u_i,k),c_i=\mbox{cn}(u_i,k),d_i=\mbox{
dn}(u_i,k) $, $i=1,2$ and formulae (\ref{b1},\ref{b2}) for the
Kleinian hyperelliptic functions.

The straightforward calculations lead to the formulae
\begin{eqnarray}
X_1=-\frac{(1-\alpha )(1-\beta )(\alpha \beta +\wp _{12})}{(\alpha
+\beta
)(\wp _{12}-\alpha \beta )+\alpha \beta \wp _{22}+\wp _{11}},
\label{x1} \cr
X_2=-\frac{(1+\alpha \beta )(\alpha \beta -\wp _{12})-\alpha \beta \wp
_{22}-\wp _{11}}{(\alpha +\beta )(\wp _{12}-\alpha \beta )+\alpha
\beta \wp
_{22}+\wp _{11}},  \label{x2} \\
X_3=\frac{\alpha \beta \wp _{22}-\wp _{11}}{(\alpha +\beta )(\wp
_{12}-\alpha \beta )+\alpha \beta \wp _{22}+\wp _{11}}.  \label{x3}\nonumber
\end{eqnarray}
The formulae (\ref{x2}) can be inverted as follows
\begin{eqnarray}
\wp_{11}=(B-1)\frac{A(X_2+X_3)-B(X_3+1)}{X_1+X_2-1},  \label{wp11} \\
\wp_{12}=(B-1)\frac{1+X_1-X_2}{X_1+X_2-1},  \label{wp12} \\
\wp_{22}=\frac{A(X_2-X_3)+B(X_3-1)}{X_1+X_2-1},  \label{wp22}
\end{eqnarray}
where $A=\alpha +\beta $, $B=1+\alpha \beta $.

The obtained results permit to present few solutions in elliptic
functions of the initial problem, which are
quasi-periodic in $\zeta$. Using (\ref{wp12}) and (\ref{wp22}) for solutions of the
(\ref{system}) in the form (\ref{solution}) we have
\begin{eqnarray*}
&&q_1^2=2\frac{1}{a_1-a_2}\left(a_1^{2}-
\frac{A(X_2-X_3)+B(X_3-1)}{X_1+X_2-1}a_1\right.\cr&&\left.\qquad-
(B-1)\frac{1+X_1-X_2}{X_1+X_2-1}\right), \\
&&q_2^2=2\frac{1}{a_2-a_1}\left(a_2^{2}-
\frac{A(X_2-X_3)+B(X_3-1)}{X_1+X_2-1}a_2\right.\cr&&\left.\qquad-
(B-1)\frac{1+X_1-X_2}{X_1+X_2-1}\right),\end{eqnarray*}
where
\begin{equation}
u_{\pm }=-2\sqrt{(1-\alpha )(1-\beta )}(x\mp c )
\end{equation}
and $c$ is the constant depending on initial conditions.

We also remark, that the derived quasi perodic
solution was associated with the Jacobi reduction case in which the
ultraelliptic integrals were reduced to elliptic ones by the aid of
second order substitution. This means on the language of two-dimensional
$\theta$-functions, that the associated period matrix is equvalent to the
matrix with the off-diagonal element $\tau_{12}=\frac12$. Such the reduction
case was considered in various places (see e.g.\cite{bbeim94}).
Solutions of this type for nonlinear Schr\"odinger equation ($\sigma=0$) are
recently obtained in \cite{ch95}.

The anologous
technique can be carried out for other well documented case of reduction ,
when $\tau_{12}=1/N$ and the $N=3,4,\ldots$. In general such the
reduction can be caried out for covers of arbitrary degree within the
Weierstrass-Poincar\'e reduction theory
(see e.g. \cite{kr03,bbeim94}).

\section{Elliptic periodic solutions}
In this section we develop a method (see also
\cite{k89,ek94,ee94b}) which allows us to construct periodic
solutions   of (\ref{system})  in a straightforward way
based on  the application  of spectral theory for the Schr\"odinger equation
with elliptic potentials \cite{amm77,mm75}.
We start with the formula (\ref{eeq1}) and with equation with equation for Baker function
$\Psi(\lambda;\boldsymbol{ u})$.
\begin{eqnarray}
&&\frac {\mathrm{d}^2} {\mathrm{d} x^2} \Psi(\lambda,\boldsymbol{ u})
-{\mathsf  U} \Psi(x,\boldsymbol{ u}) =
(\lambda+\frac{\alpha_{4}}{4})\Psi(\lambda,\boldsymbol{ u}),  \label{baker}
\end{eqnarray}
where we identify the potential $${\mathsf  U}=2\wp_{22}+\frac16\alpha_{4}.$$
We assume, without loosing generality, that the
associated curve has the property $\alpha_4=0$. To make this assumption
applicable to the initial curve of the system (\ref{system}) being derived
from the Lax representation, we undertake the shift of the spectral parameter,
\begin{equation}
\lambda\longrightarrow \lambda+\Delta,\qquad \Delta=\frac25a_1+\frac25a_2.
\label{shift}
\end{equation}

Suppose, that ${\mathsf  U}$ be two gap Lam\'e or two
gap {\it Treibich-Verdier} potential, what means,
that
\begin{equation}
{\mathsf  U}(x)=2\sum_{i=1}^N \wp(x-x_i) \label{TVP},
\end{equation}
where $\wp(x)$ is standard Weierstrass elliptic functions with periods $2\omega,2\omega'$ and numbers
$x_i$ takes the values from the set $\{ 0,\omega_1=\omega,\omega_2=\omega+\omega',\omega_3=\omega'\}$.
It is known, that the set of such the potentials is exhausted
by six potentials\cite{tv90,ee95a}
\begin{eqnarray}
{\mathsf  U}_3(x)&=&6\wp(x), \label{L3} \\
{\mathsf  U}_4(x)&=&6\wp(x)+2\wp(x+\omega_i),\quad i=1,2,3,\label{TV4}\\
{\mathsf  U}_5(x)&=&6\wp(x)+2\wp(x+\omega_i)+2\wp(x+\omega_j),
\cr&& \qquad\qquad i\neq j=1,2,3,
\label{TV5}\\
{\mathsf  U}_6(x)&=&6\wp(x)+6\wp(x+\omega_i),\quad i=1,2,3,\nonumber\\
{\mathsf  U}_8(x)&=&6\wp(x)+2\sum_{i=1}^3\wp(x+\omega_i),
\nonumber\\
{\mathsf  U}_{12}(x)&=&6\wp(x)+6\sum_{i=1}^3\wp(x+\omega_i),
\nonumber
\end{eqnarray}
where the subscript shows the number of $2\wp$ functions involved and
display the degree of the cover of the associated genus two curve over
elliptic curve. Because the last three potentials can be obtained from the
first three by Gauss transform we shall call the first three as
{\it basis potentials}. The potential (\ref{L3}) is two gap  Lam\'e
potential, which is associated with three sheeted cover of elliptic curve;
the potentials (\ref{TV4},\ref{TV5}) are Treibich-Verdier potentials
\cite{ve90,tv90} associated with four and five sheeted cover correspondingly.

To display the class of periodic solutions of system (\ref{system})
we introduce the {\it generalized Hermite polynomial}
${\mathcal  F}(x,\lambda)$ by the formula
\begin{equation}
{\mathcal  F}(x,\lambda)=\lambda^2-\pi_{22}(x)\lambda-\pi_{12}(x)
\end{equation}
with $\pi_{22}(x)$ and $\pi_{12}(x)$ given as follows
\begin{eqnarray}
\pi_{22}(x) &=&\sum_{j=1}^N \wp (x - x_j) +
\frac13 \sum_{j=1}^5 \lambda_j,\label{tr1} \cr
\pi_{12}(x) &=&-3\,\sum_{i<j} \wp(x - x_i) \wp(x -x_j) -
\frac{Ng_2}{8} \nonumber\\
&&-\frac16 \sum_{i<j} \lambda_i \lambda_j + \frac{1}{
6}\left(\sum_{j=1}^5 \lambda_j^2 \right) \label{tr2}
\end{eqnarray}
where $x_i$ are half-periods and $N$
is the degree of the cover (see for example \cite{ek94}). The introduction
of this formula is based on the possibility
to compute the symmetric function $\mu_1\mu_2$ in terms of differential
polynomial of the first one with the help of
the equation (\ref{eeq1}), which serves in this context as the
``trace formula" \cite{zmnp80}.

The solutions of the system (\ref{system}) are then given as
\begin{equation}
q_1^2(x)=2\frac{{\mathcal  F}(x,a_{1}-\Delta)}
{a_{1}-a_{2}} , \quad
q_2^2(x)=2\frac{{\mathcal  F}(x,a_{2}-\Delta)}
{a_{2}-a_{1}} ,\label{answer}
\end{equation}
 The final formula for
the solutions of the system (\ref{manakov}) then reads

\begin{eqnarray}
{\mathcal  U}(x,t)=\,\sqrt{2\frac{{\mathcal  F}(x,a_{1}-\Delta)}
{a_{1}-a_{2}}}
\mathrm{ exp}\left \{\imath a_1 t-\frac12\nu(a_i-\Delta)\int\limits_{\cdot}^x
 \frac{{\mathrm d}x}{{\mathcal F}(x,a_1-\Delta)}\right\},\cr
 \label{final}\\
{\mathcal  V}(x,t)=\sqrt{2\frac{{\mathcal  F}(x,a_{2}-\Delta)}
{a_{2}-a_{1}}}
\mathrm{ exp}\left \{\imath a_2 t-\frac12\nu(a_2-\Delta)\int\limits_{\cdot}^x
 \frac{{\mathrm d}x}{{\mathcal F}(x,a_2-\Delta)}\right\}
,\nonumber
\end{eqnarray}
where we used (\ref{answer}) and (\ref{ccc}).

It is important to remark for our consideration, that if the potential
is known, then the associated algebraic curve of genus two can be
described with the help of the Novikov equation \cite{no74}.  Let us consider
the two-gap potential for normalized by its expansion near the singular point as
\begin{equation}
{\mathsf  U}(x) = \frac{6}{ x^2} + a x^2 + b
x^4 + c x^6 + d x^8 + O(x^{10}).  \label{decomposition}
\end{equation}
Then the algebraic curve associated with this potential has the
form \cite{be89b}
\begin{eqnarray}
\nu^2&=&\lambda^5 - \frac{5\cdot7}{2} a\lambda^3
+\frac{3^2\cdot7}{2} b\lambda^2 \cr
&+& \left( \frac{3^4\cdot7}{8} a^2 +\frac{3^3\cdot11}{4}c  \right)\lambda
 -\frac{3^4\cdot17}{4}ab+\frac{3^2\cdot11\cdot13}{2}d\label{curvelame}.
\end{eqnarray}

We shall consider below examples of genus two curves, which are associated
with the two gap elliptic potentials (\ref{L3}), (\ref{TV4}) and (\ref{TV5}).

Consider the potential ${\mathsf U}_3$ and construct the associated curve
(\ref{curvelame})
\begin{equation}{\mathcal L}^2=(\lambda^2-3g_2)(\lambda+3e_1)
(\lambda+3e_2)(\lambda+3e_3),\label{curve3}\end{equation}
 The Hermite polynomiaal ${\mathcal F}_3(\wp(x),\lambda)$
\cite{ww86} associated to the Lame potential (\ref{L3}), which is already
normalized as in (\ref{decomposition}) has the form
\begin{equation}
{\mathcal F}_3(\wp(x),\lambda)=\lambda^{2}-
3\wp(x)\lambda +
9\wp^{2}(x)-\frac{9}{4}g_{2}.  \label{HerPol}
\end{equation}
Then the finite and real solution to the system (\ref{system})
is given by the formula (\ref{answer}) with the Hermite polynomial
depending in the argument $x+\omega'$ (the shift in $\omega'$ provides
the holomorphity of the solution). The solution is real under the choice
of the arbitrary constants $a_{1,2}$ in such way, that the constants
$a_{1,2}-\Delta$  lie in {\it different } lacunae.
According to (\ref{ccc}) the constants $C_{i}$ are
then given as
\[C_i^2=-\frac{4\nu^2(a_i-\Delta)}{(a_i-a_j)^2}\label{cccc},
\]
where $\Delta$ is the shift (\ref{shift}) and $\nu$ is the coordinate
of the curve (\ref{curve3}) and the integrals $H$ and $F$ have the
following form
\begin{eqnarray}
&& H=\frac{1}{25}\left(a_1+a_2\right)^3+\frac{21}{4} g_2, \nonumber \\
&& F=\frac{1}{25}\left(a_{{1}}+a_{{2}}\right)^{3}-\frac{1}{4}C_{1}^{2}
-\frac{1}{4}C_{2}^{2}-\frac{27}{4}g_{{3}}-\frac{21}{20}g_{{2}}
\left(a_{1}+a_{2}\right). \nonumber
\end{eqnarray}

These results are in complete agreement with solutions obtained
in \cite{pp99} by introducing the ansatz of the form
$$q_i(x)=\sqrt{A_i\wp(x)^2+B_i\wp(x)+C_i},\quad i=1,2$$ with the constants
$A_i,B_i,C_i$ which are defined from the compatibility condition of
the ansatz with the equations of motion. In the foregoing examples we are
considering the solutions of the form
$$q_i(x)=\sqrt{{\mathcal Q}_i(\wp(x))}, $$
where ${\mathcal Q}_i$ are rational functions of $\wp(x)$.

Consider with this purpose the Treibich Verdier potential
\begin{equation}
{\mathsf  U}_4(x)=6\wp(x)+2\wp(x+\omega_1)-2e_1,\label{TV4}
\end{equation}
associated with four sheeted cover. The potential is normalized according
to (\ref{decomposition}). The associated spectral curve is of the form

\begin{eqnarray}{\nu}^2&=&4(\lambda+6e_1)\prod_{k=1}^4
(\lambda-\lambda_k)
\label{ctv4}\\
\lambda_{1,2}&=&e_3+2e_2\pm 2\sqrt{(5e_3+7e_2)(2e_3+e_2)}\\
\lambda_{3,4}&=&e_2+2e_3\pm 2\sqrt{(5e_2+7e_3)(2e_2+e_3)}\nonumber,
\end{eqnarray}
The Hermite polynomials are  given by the formula
\begin{eqnarray} {\mathcal F}(x,\lambda)&=&
\lambda^2-(3\wp(x)+\wp(x+\omega_1)-e_1)\lambda\label{hertv4}\\&+&
9\wp(x)(\wp(x)+\wp(x+\omega)-e_1)-3e_1\wp(x+\omega_1)\cr&+&
\frac{9}{4}g_2-51e_1^2;
\nonumber
\end{eqnarray}
The finite real solution of (\ref{system}) results the
substitution this Hermite polynomial
$ {\mathcal F}(x+\omega',\lambda)$  into (\ref{answer}) depending in shifted
by imaginary half period argument into the formula (answer). To provide
the reality of the solution we shall fix the parameters $a_i-\Delta$
in the permitted zones. The constants $C_i$ are computed by the formula
(\ref{cccc}) at which $\nu$ means the coordinate of the curve (\ref{ctv4}).

Consider further the Treibich Verdier potential
\begin{equation}
{\mathsf  U}_5(x)=6\wp(x)+2\wp(x+\omega_2)+2\wp(x+\omega_3)+2e_1,
\label{TV4}
\end{equation}
associated with four sheeted cover. The pontial is normalized according to
(\ref{decomposition}). The associated spectral curve is of the form

\begin{eqnarray}\nu^2&=&(\lambda+6e_2-3e_3)(\lambda+6e_3-3e_2)
\nonumber\\
&\times&\left[\lambda^3+3e_1\lambda^2-(29e_2^2-22e_2e_3+29e_3^2)\lambda
\right.\cr
&+&\left.159(e_2^3+e_3^3)-51e_2e_3(e_2+e_3)\right] \label{ctv5}
\end{eqnarray}
The associated Hermite polynomials are given by the formula
\begin{eqnarray*} {\mathcal F}(x,\lambda)
&=&\lambda^2-(3\wp(x)+\wp(x+\omega_2))+\wp(x+\omega_3)+e_1)
\lambda\\
&+&9\wp(x)(\wp(x)+\wp(x+\omega_2)+\wp(x+\omega_3
) )+3\wp(x+\omega_2)\wp(x+\omega_3)\cr&+&3e_1(3\wp(x)+\wp(x+\omega_2))
+\wp(x+\omega_3))-\frac{39}{2}g_2+54e_1^2.
\end{eqnarray*}
The solution of the system results the substitution of
these expressions to (\ref{answer}) as before, but this solution is
blowing up.

We remark, that since McKean, Moser and Airlault paper \cite{amm77} is well known, that all
elliptic potentials of the Schr\"odinger equations and their isospectral
transformation under the action of the KdV flow has the form,
\begin{equation}
{\mathsf  U}(x)=2\sum_{i=1}^N\wp(x-x_i(t)),\label{elpot}
\end{equation}
The number $N$ is a positive integer $N>2$ (the
number of ``particles") and the numbers $\boldsymbol{ x}=(x_1(t),\ldots,x_N(t))$
belongs to the locus ${\mathcal  L}_N$, i.e., the
geometrical position of the points given by the equations
\begin{equation}
{\mathcal  L}_N=\left\{(\boldsymbol{ x});\sum_{i\neq
j}\wp'(x_i(t)-x_j(t))=0,\; j=1,\ldots N\right\}.\label{locus}
\end{equation}
If the evolution of the particles $x_i$ over the locus is given by the
equations,
\[\frac{{\mathrm d} x_i}{ {\mathrm d} t}=6\sum_{j\neq i}\wp(x_i(t)-x_j(t))
\]
then the potential (\ref{elpot}) is the elliptic solution to the KdV equation.
Henseforth the elliptic potentials discuss can serve as imput for the isospectral deformation
along the locus. Moreover these elliptic potential do not exhausted all the viriety of elliptic
potential; we can mention here the elliptic potentials of Smirnov
(\cite{sm89,sm94}) for which the shifts
$x_i$ are not half periods. The involving of these objects to the
subject can enlarge the classes of
elliptic solutions to the system  (\ref{manakov})

\section{Conclusions}

In this paper we have described a family of elliptic solutions for the coupled
nonlinear Schr\"dinger equations using Lax pair method and the general method
of reduction of Abelian functions to elliptic functions. Our approach is
systematic in the sense that special solutions (periodic, soliton etc.) are
obtained in a unified way. We considered only the family of elliptic solutions
associated with the integrable case $1:2:1$ of quartic potential, the approach
developed can be applied to other integrable cases being enumerated in the
introduction.

In fiber optics applications of the periodic and quasi-periodic  waves are
of interest in optical transmission systems.

\end{document}